\documentclass[manuscript=article,journal=jcim]{achemso}
\usepackage{graphicx} 
\usepackage{subcaption}
\usepackage{hyperref}
\usepackage{amsmath}
\usepackage{booktabs}
\usepackage{achemso}

 \title{Graph Neural Networks for Carbon Dioxide Adsorption Prediction in Aluminium-Exchanged Zeolites}
\author{Marko Petkovi\'{c}}
\author{Jos\'{e} Manuel Vicent-Luna}
\author{Vlado Menkovski}
\author{Sof\'{i}a Calero}
\email{s.calero@tue.nl}
\affiliation{Eindhoven University of Technology, 5612AZ Eindhoven, Netherlands.}

\begin{document}

\maketitle

\begin{abstract}
The ability to efficiently predict adsorption properties of zeolites can be of large benefit in accelerating the design process of novel materials. The existing configuration space for these materials is wide, while existing molecular simulation methods are computationally expensive. In this work, we propose a model which is 4 to 5 orders of magnitude faster at adsorption properties compared to molecular simulations. To validate the model, we generated datasets containing various aluminium configurations for the MOR, MFI, RHO and ITW zeolites along with their heat of adsorptions and Henry coefficients for CO$_2$, obtained from Monte Carlo simulations. The predictions obtained from the Machine Learning model are in agreement with the values obtained from the Monte Carlo simulations, confirming that the model can be used for property prediction. Furthermore, we show that the model can be used for identifying adsorption sites. Finally, we evaluate the capability of our model for generating novel zeolite configurations by using it in combination with a genetic algorithm.
\end{abstract}

\section{Introduction}
Over the past years, the amount of CO$_2$ in the atmosphere has been increasing, and greenhouse effects have become increasingly evident. A potential way to reduce the carbon levels in the atmosphere is carbon capture \cite{odunlami2022advanced}. Nanoporous materials, such as zeolites, are good candidates \cite{kumar2020utilization}, due to their high adsorption capacity of gasses \cite{gargiulo2014co2,gutierrez2011molecular,gutierrez2013molecular}. In addition, zeolites have a high thermal stability \cite{cruciani2006zeolites}, and their synthesis can have a low cost \cite{khaleque2020zeolite} compared to other adsorbents. Furthermore, there exists a large number of synthesizable zeolite topologies \cite{derbe2021short}, with different pore sizes and properties. For each zeolite topology, there exist multiple possible configurations depending on the amount of aluminium and silicon atoms. Here, the Si/Al ratio can influence the properties of the zeolite \cite{choi2022effect,moradi2021effect,yang2018atomistic}, such as the adsorption of CO$_2$. While the overall trend shows that decreasing the Si/Al ratio leads to a higher heat of adsorption, there can still be a lot of variance in the heat of adsorption for a single framework with a set Si/Al ratio \cite{romero2023adsorption}. 

As a result of the amount of possible Si/Al configurations, the space of possible zeolites is very large, such that it is not feasible to experimentally study the different configurations. An alternative is using molecular simulations, which have proven to be an excellent tool to understand and predict material properties \cite{garcia2007computational,broclawik2021zeolites,smit2008molecular,gutierrez2013molecular,luna2015understanding}. However, carrying out a simulation for a single property for a single configuration can still take hours to days. Combined with the large search space, obtaining the properties of all different configurations using molecular simulations is still unfeasible. Recently, Machine Learning (ML) models have been developed for property prediction of molecules and materials \cite{choudhary2022recent,stein2019progress,reiser2022graph}. They have shown excellent performance, while being able to make predictions almost instantaneously.  

Existing ML methods for property prediction of zeolites and other nanoporous materials often rely on hand crafted descriptors and traditional ML algorithms such as Random Forests (RF), Support Vector Machines (SVM) and Multi Layer Perceptrons (MLP) \cite{fanourgakis2019robust,fanourgakis2020generic,burner2020high,anderson2020adsorption}. These descriptors usually include features like surface area, largest pore diameter and pore limiting diameter. Other methods create descriptors from the geometry of the materials, using topological data analysis techniques such as persistent homology \cite{zhang2019machine,krishnapriyan2020topological}. However, these descriptors may struggle to capture the relationship between the crystal structure and the target property. As such, the model's performance is strongly dependent on the quality of the descriptors. Furthermore, these descriptors often remain the same for a particular zeolite topology, regardless of the Si/Al ratio. One potential direction to address these limitations is using end-to-end Deep Learning (DL) models, which take the material structure as input and produce predictions in one step. DL models make use of representation learning, meaning that the model learns which features are relevant to its predictions, instead of using handcrafted features. 


Another advantage of DL approaches is their ability to be in- and equi-variant to various symmetries, such as translations, rotations and reflections. In turn, by incorporating these symmetries in a model, it can become more efficient with regards to the amount of training data, as well as demonstrate better generalization performance. Convolutional Neural Networks (CNN) consist of convolutional layers that are equivariant to translations, allowing the network can detect the same pattern in various locations in an image. \citet{lu2022deep} proposed a method based on CNNs, where the atoms are first encoded in three different matrices based on their x-y, y-z and x-z coordinates. Then, the three channel image is processed with a ResNet \cite{he2016deep} to obtain predictions. In this data representation, a part of the three dimensional geometric structure is missing, which could impact performance. One can overcome this limitation by using a 3D-CNN to predict material properties \cite{cho2021nanoporous}. In their dataset, the structure of the zeolite is represented as a three dimensional grid, where pixels take values based on the availability for adsorption of that site. The grid is calculated by probing each coordinate in the zeolite for availability \cite{willems2012algorithms}. However, the model cannot directly take different atoms into account (Si/Al). In addition, the representation does not explicitly take into account the effects of cations on the availability of adsorption sites. It also does not make use of the periodic structure of zeolites, which can limit its expressivity, as part of the model parameters need to be used to learn equivalences of symmetry groups.

Some of these challenges can be addressed using Graph Neural Networks (GNNs). Materials can be represented using atoms as nodes, while edges could be drawn to represent covalent bonds, or based on distance. As such, this representation can be calculated directly from reading the CIF of a crystal. Different GNNs such as Crystal Graph CNN (CGCNN) \cite{xie2018crystal}, ALIGNN \cite{choudhary2021atomistic} and Schnet \cite{schutt2017schnet} have been successfully applied for predicting material properties. These models take the periodic structure of crystalline materials into account and are also equivariant/invariant to transformations from the Euclidean Group (E(3)).

For example, CGCNN has been used to accelerate methane adsorption hierarchical screening \cite{wang2020accelerating}, where each atom in a MOF is represented by a node. In a similar approach using CGCNN \cite{wang2022combining}, Secondary Building Units (SBU) of MOFs are used as nodes instead. The drawback of the graph representation of materials, is that the geometry (in terms of pores) is not explicitly encoded in the representation, since the empty space at the adsorption site is not explicitly defined in a graph.

\begin{figure}[ht!]
    \centering
    \includegraphics[width=.8\textwidth]{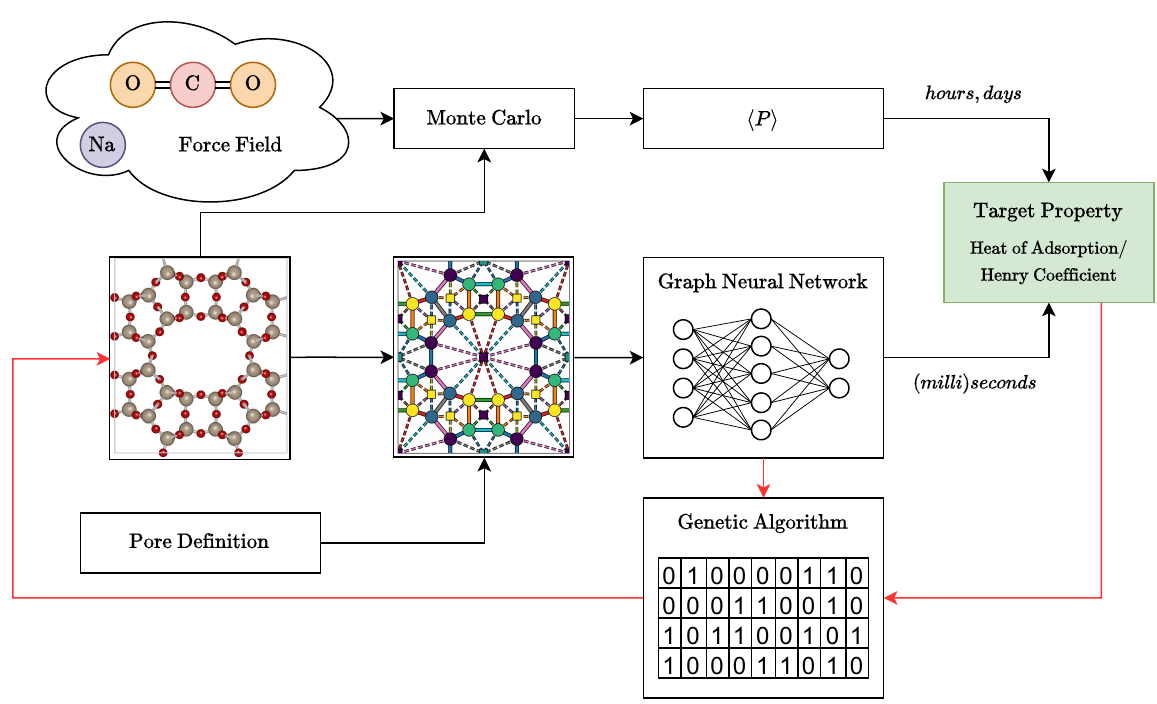}
    \caption{Pipeline comparison between calculating heat of adsorption using Monte Carlo simulations and our proposed ML method. Black arrows represent the property prediction process, while red arrows represent the zeolite generation process.}
    \label{fig:pipeline}
\end{figure}

For the GNN to form the notion of a pore, enough message passing steps are needed, such that all atoms surrounding a pore have exchanged information with each other. However, zeolites can have large pores (such as the 12-ring in MOR), which require many message passing steps to connect each atom. In turn, this can lead to over-smoothing, where node representations become very similar, leading to a loss of discriminative information in the graph. As such, a GNN might struggle to implicitly learn the strength of different adsorption sites in a material. To overcome this limitation, the Equivariant Porous Crystal Networks (EPCN) \cite{petkovic2023equivariant} architecture was proposed, which a GNN model for heat of adsorption prediction, where the porous structure of zeolites is explicitly encoded in the graph. This was achieved through the inclusion of pore nodes, which were placed in the pores of the material and were connected with the atoms surrounding that pore. As a result, the network obtained a higher performance than other GNN approaches which do not explicitly encode the pores. Furthermore, this method additionally exploits the symmetry group of zeolites, by sharing parameters between equivalent nodes/edges, increasing the expressivity of the method. By training such a method, the adsorption properties of new zeolites can be calculated more rapidly (almost instantly), compared to using traditional methods, such as Monte Carlo, where simulating a single zeolite can take hours to days (Figure \ref{fig:pipeline}). In EPCN, the final hidden state of each pore is further processed with an MLP after aggregation. In turn, it is not possible to find a direct relationship between the heat of adsorption and the various pores. Thus, we cannot use the model to understand where in the zeolite adsorption happens.

In this work, we extend the EPCN model \cite{petkovic2023equivariant} in such that it allows us to identify the adsorption sites, without a significant change in performance. This makes the model interpretable, as we can now identify the pores responsible for the adsorption. Furthermore, we extend the dataset provided with ECPN (CO$_2$ heat of adsorption for MOR and MFI) with CO$_2$ heat of adsorption values for different configurations of the RHO and ITW zeolites, as well as with Henry coefficients for all four zeolites. In addition, we extend the model to predict the Henry coefficient of each zeolite configuration in the dataset. Using this dataset, we show that the model performs excellently predicting the CO$_2$ heat of adsorption and Henry coefficient simultaneously, while being significantly faster than Monte Carlo simulations. Finally, we demonstrate that this model can be used for inverse design of zeolites based on their heat of adsorption, by using the model for the fitness function of a genetic algorithm.

\section{Methods}
\subsection{Zeolite Frameworks}
For this work, we made use of the MOR, RHO, MFI and ITW zeolite topologies. For each topology, we generated different configurations of aluminium and silicon atoms, varying between 0 to 12 for MOR, RHO and ITW, and between 0 and 24 for MFI. The configurations were generated using the ZEORAN program \cite{romero2023adsorption}. The program can generate configurations of zeolite topologies with a given amount of aluminium atoms. Depending on the algorithm used for generating the structures, the generated structures might break the L{\"o}wenstein rule \cite{loewenstein1954distribution}, which prohibits bonding of two aluminium atoms through an oxygen atom (Al-O-Al). Since multiple recent studies \cite{afeworki2004synthesis,heard2019effect,pavon2014direct,fletcher2017violations} found violations of the L{\"o}wenstein rule in zeolites, structures which break the rule were also included in the study.

The ZEORAN program can generate Al/Si configurations for a zeolite using four algorithms. The \textit{cluster} algorithm places aluminium atoms clustered together in the zeolite topology. In the \textit{chains} algorithm, aluminium atoms are placed in such a way that they form chains in the zeolite. In this algorithm, the length of each chain is defined by the user. The \textit{maximum entropy} algorithm places the aluminium atoms such that they are evenly distributed through the zeolite. The final algorithm places aluminium atoms \textit{randomly} throughout the zeolite. For each amount of aluminium substitutions in a zeolite, each algorithm generated roughly one quarter of the structures. More details on the zeolite frameworks can be found in Table \ref{tab:frameworks}.

\begin{table}[t!]
    \centering
    \caption{Properties of different zeolite frameworks}
    \vspace{2mm}
    \begin{tabular}{lcccc}
\toprule
Framework &                  \# structures & \# atoms &                  Min. Si/Al ratio  & Sym. group size \\
\midrule
MOR & 4992 & 48 & 3.0 & 16 \\
MFI & 3296 & 96 & 3.0 & 8 \\
RHO & 1212 & 48 & 3.0 & 48 \\
ITW & 762 & 24 & 1.0 & 8 \\
\bottomrule
\end{tabular}
    \label{tab:frameworks}

\end{table}

\subsection{Computational Details}
In this study, we investigated the heat of adsorption ($-\Delta H$) and Henry coefficient ($K_H$) \cite{do2008henry, prasetyo2018coherent}, which reflect the interaction strength between the CO$_2$ and the zeolite. To calculate these two properties, Monte Carlo (MC) simulations using the Widom particle insertion method in the canonical ensemble (NVT) were performed \cite{widom1963some}. To obtain the heat of adsorption value, Equation \ref{eq:hoa} was used, where $\Delta U$ represents the internal energy difference before and after the adsorption of the guest molecule. In this equation, we neglect the contributions to the internal energy of the zeolite and the CO$_2$ molecules, since they are modeled as rigid bodies.    
\begin{equation}
-\Delta H = \Delta U - RT
    \label{eq:hoa}
\end{equation}
The Henry coefficient is the constant that relates the number of adsorbed molecules per unit of volume ($\theta$) and the external pressure in the infinite dilution regime ($\theta=K_HP$). This coefficient can be calculated using the Widom insertion method as:  

\begin{equation}
K_H = \frac{1}{N} \beta \sum_{i=1}^N e^{-\beta \Delta U_i}
    \label{eq:henry}
\end{equation}
\noindent where $\beta=\frac{1}{k_BT}$, being $k_B$ the Boltzmann constant, and N is the number of configurations.

Atomic coordinates for all pure silica zeolites were taken from the IZA database \cite{baerlocher2007atlas}, following which configurations with variable aluminium substitutions were created using the ZEORAN program \cite{romero2023adsorption}. Additional sodium cations were added in each simulation box to compensate for the change in charge as a result of substituting silicon with aluminium. All simulations were carried out using the RASPA software \cite{dubbeldam2016raspa}. The same simulation settings, force field and point charges as in \citet{romero2023adsorption} were used, which extends the force field and point charges from \citet{garcia2009transferable}, to model the atoms that break the L{\"o}wenstein rule.   

\subsection{Dataset}

\begin{figure}[hb!]
    \centering
    \includegraphics{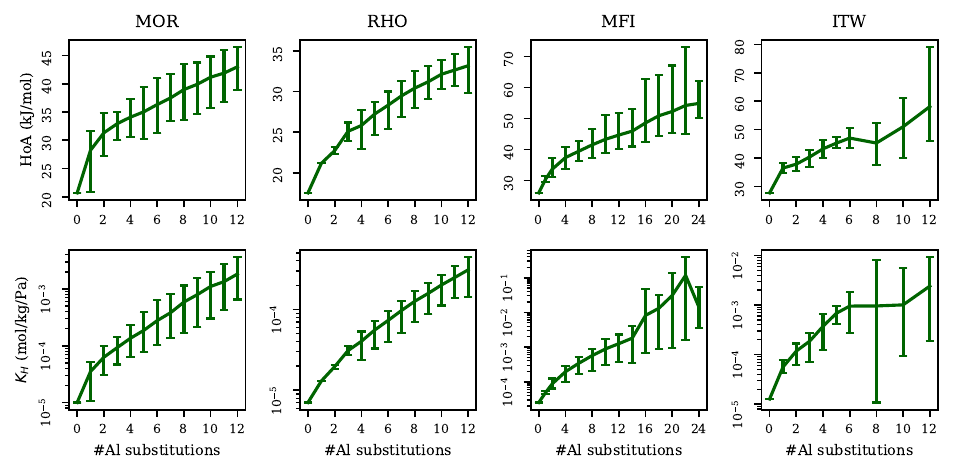}
    \caption{Heat of adsorption and Henry coefficient distribution for different topologies. Vertical bars indicate the 95\% confidence interval for each amount of aluminium substitutions substitutions.}
    \label{fig:dataset}
\end{figure}

In Figure \ref{fig:dataset}, we visualised the distribution of the heat of adsorption and Henry coefficient obtained from MC simulations for each zeolite topology. Overall, both the heat of adsorption and Henry coefficient seem to increase with the number of aluminium atoms inside the structure. However, there is still a big variance in both properties for zeolites with a set number of aluminium atoms. As has been shown by \citet{romero2023adsorption}, sodium cations are typically found close to the aluminium framework atoms. In turn, the distribution of the aluminium atoms in the framework affects the location and strength of adsorption sites. In addition, there is a big overlap between heat of adsorption and Henry coefficient distributions between configurations with different numbers of aluminium atoms. Therefore, more complicated models are necessary to model these properties based on the zeolite topology and composition.
\subsection{Models}
\begin{figure}[hb!]
    \centering
    \includegraphics{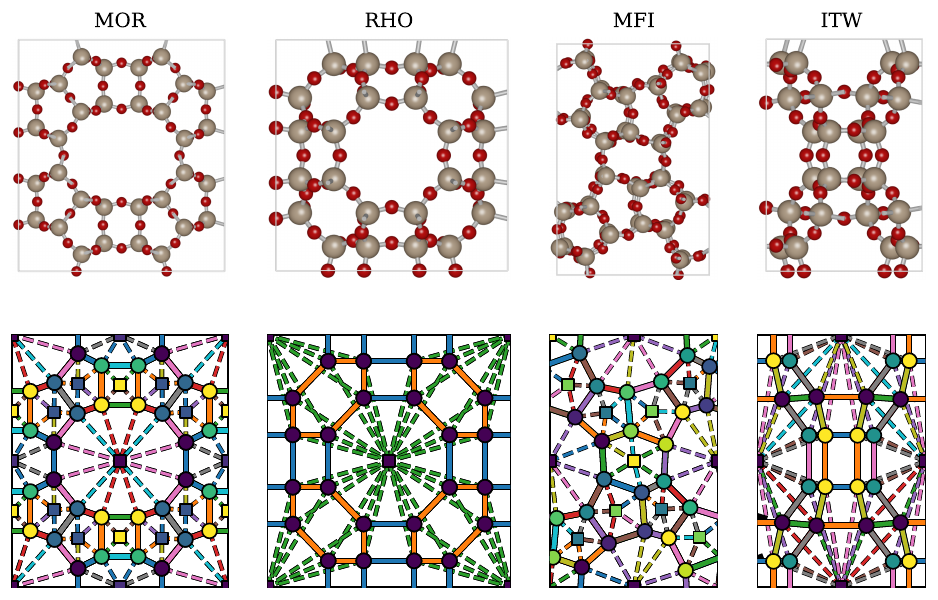}
    \caption{Zeolite structures used in this work. MOR, RHO and ITW are visualised along the z-axis, while MFI is visualised along the y-axis. Top row: Zeolite structures visualised using iRASPA \cite{dubbeldam2018iraspa}. Bottom row: Zeolite graph representation for ML. Circles and squares represent T-atom nodes and pore nodes, while solid edges are drawn between T-atoms and dotted edges between T-atoms and pores. Edges/Nodes of the same color and type are symmetric, and thus share parameters. Note that for MFI we only visualised the top layer of atoms.}
    \label{fig:zeovsml}
\end{figure}

To model the heat of adsorption of the different zeolites, we make use of Graph Neural Networks (GNNs). To encode a zeolite as a graph, we represent the T-atoms (Al/Si) as nodes in the graph. In the graph, edges are drawn between T-atoms which share an oxygen atom, as well as between T-atoms and the corresponding pore nodes. We explicitly omit encoding the oxygen atoms in the graph, as these always remain in the same position within a framework, while the T-atoms can change. However, one could say that the oxygen atoms are implicitly represented by the edges of the graph, as these are drawn between atoms connected by an oxygen. Atom nodes are represented by 1 if they are aluminium and 0 if they are silicon, while for pore nodes the area and the ring size of the pore are part of the feature vector. On each edge, the distance between two T-atoms is encoded using radial basis functions (Equation \ref{eq:rbf}). In these functions, $\gamma$ and $\boldsymbol{\mu}$ are hyperparameters, while $\mathbf{x}_i$ and $\mathbf{x}_j$ are the nodes between which the distance is calculated. When calculating the distance between two nodes, we take the periodic boundary conditions into account using the minimum image convention. 
\begin{equation}\label{eq:rbf}
    \mathbf{e}_{ij} = \exp \left(-\gamma(\|\mathbf{x}_i -\mathbf{x}_j\| - \boldsymbol{\mu})^2 \right)
\end{equation}
To efficiently model the porous structure of the material, we made use of the EPCN architecture \cite{petkovic2023equivariant}. This architecture extends message passing neural networks \cite{gilmer2017neural}, by making use of the symmetries present in crystalline materials \cite{kaba2022equivariant}. Here, edges and nodes which can be mapped onto each other using transformations from the symmetry group of the zeolite share a unique set of parameters. The parameter sharing scheme for the zeolites we used can be found in Figure \ref{fig:zeovsml}, where nodes/edges with the same color share parameters. Since the parameter sharing is invariant to the space group of the zeolites, the network layers remain equivariant, while being more expressive than a regular GNN. 

In the EPCN architecture, the final hidden states of each pore are aggregated through feature-wise sum-pooling. Following this aggregation, the aggregated hidden state is processed by an MLP to make a (scalar) prediction. In our experiments, we remove the aggregation over the hidden states of the pore, and instead process the hidden states of each pore by the same MLP to obtain a scalar value for each property. We then sum the scalar values for each pore to obtain the output of the network. As such, the network is implicitly forced to learn the contribution of each pore to the adsorption properties of the zeolite, resulting in an explainable model. 

\subsection{Model Training}
In our experiments, we randomly assigned samples from each zeolite to a training and testing set, where each testing set consists of 10\% of the datapoints from a zeolite. This was done to reflect practical uses of such models, where one could first simulate enough materials covering the whole range of aluminium substitutions, before training a model for predicting the properties of new configurations. 

We trained EPCN along with our extension to model the different zeolites. For both networks, the hidden states have a size of 8, while the final hidden state has a size of 24. The main difference between the two networks is that the original network performs sum-pooling on the final hidden state of each pore, while our extension performs the sum-pooling on the output of the network.

All models were trained for 200 epochs, using the AdamW optimizer \cite{loshchilov2018decoupled} with a learning rate of 0.001 and a batch size of 32. We evaluate the model performance using the Mean Absolute Error (MAE), Mean Squared Error (MSE) and the coefficient of determination ($R^2$). The MAE measures the average error (Equation \ref{eq:mae}) between the predicted ($\hat{y}_i$) and true value ($y_i$), while the MSE measures the average squared error (Equation \ref{eq:mse}), and thus gives a higher weight to bigger differences between the prediction and true value. Finally, $R^2$ can be seen as a measure of how much of the variance in the target variable can be explained by the model (Equation \ref{eq:r2}), compared to a simple model which predicts the mean ($\overline{y}$). To obtain 95\% confidence intervals for the model performance indicators, we trained each model ten times. Each time, we performed a random weight initialization.  
\begin{align}
    & \text{MAE} = \frac{1}{n} \sum_{i=0}^n | \hat{y}_i - y_i | \label{eq:mae}\\ 
    & \text{MSE} = \frac{1}{n} \sum_{i=0}^n  (\hat{y}_i - y_i)^2 \label{eq:mse}\\
    & R^2 = 1 - \frac{\sum_{i=0}^n  (\hat{y}_i - y_i)^2}{\sum_{i=0}^n  (\overline{y} - y_i)^2} \label{eq:r2}
\end{align}

\subsection{Inverse Design}
In addition to predicting properties of zeolites, our model can also be used for inverse design of zeolites. This is achieved using a genetic algorithm which can generate zeolites satisfying a target heat of adsorption. In this algorithm, the binary genes directly represent the configuration of a zeolite, where 1 indicates an aluminium atom and 0 indicates a silicon atom. Each gene directly corresponds to an atomic position in the zeolite. To optimize the structures, we make use of the fitness function shown in Equation \ref{eq:fitness}, which calculates the squared error between the target heat of adsorption and current heat of adsorption predicted by the model, as well as the amount of aluminium atoms in the solution. Here, $x$ is the binary atom vector, $f_\theta$ the is trained model, $y$ is the target heat of adsorption and $\beta$ is a weighting factor. In our experiments, we set $\beta$ to 1. It can be set to a higher value if configurations with a smaller amount of aluminium atoms are desired, or to a lower value in case the amount of aluminium atoms is less not important. Likewise, the fitness function can be modified with other conditions, such as requiring a certain amount of aluminium atoms.
\begin{equation}
    f(x) = -(f_\theta(x) - y)^2 - \beta\sum(x)
    \label{eq:fitness}
\end{equation}

To implement the genetic algorithm, we used the \texttt{PyGAD} \cite{gad2021pygad} library. In the algorithm, we use 50 generations, with 2 parents mating at each generation. The crossover type is set to single point with a probability of 0.2. At each generation, 5 of the best solutions were kept. The initial population has a size of 32 and is initialized based on the target heat of adsorption. We sample the initial amount of aluminium atoms based on the distribution of the training set. This is done such that there exist structures with the same amount of aluminium atoms satisfying the target heat of adsorption. As such, some of the initial configurations should already be close to the target heat of adsorption.

Using the aforementioned settings, we generated 10 structures for each heat of adsorption from 30 kJ/mol to 55 kJ/mol for the MOR zeolite (260 structures in total). For MFI, we generated structures with heat of adsorption between 40 and 60 kJ/mol (310 structures in total). These values were chosen such that they cover the range of heat of adsorption, which contains most of the structures in the training set. Following this, we simulated the generated structures using MC.

\section{Results}
\subsection{Model Performance}
In Tables \ref{tab:results_hoa} and \ref{tab:results_henry}, the performance of the two models with respect to heat of adsorption and Henry coefficient prediction is given. For each task, the models achieve comparable performances. In all cases, no significant differences can be found, as the confidence intervals for all metrics are overlapping between the two models. As a result, we can conclude that our proposed model does not lead to a significant change in performance.

\begin{table}[t!]
    \centering
    \caption{Performance of EPCN compared to our extension on different zeolites for heat of adsorption. Bold numbers indicate the best performance for a zeolite topology.}
\small
    \begin{tabular}{lcccccc}
\toprule
{} &  \multicolumn{2}{c}{MAE $\downarrow$} & \multicolumn{2}{c}{MSE $\downarrow$} & \multicolumn{2}{c}{R$^2$ $\uparrow$}\\
\cmidrule(lr){2-3}
\cmidrule(lr){4-5}
\cmidrule(lr){6-7}
{} &                  EPCN & this work & EPCN & this work & EPCN & this work\\
\midrule
MOR & 0.89 $\pm$ 0.07 & $\mathbf{0.86 \pm 0.02}$ & 1.42 $\pm$ 0.21 & $\mathbf{1.36 \pm 0.06}$  & $\mathbf{0.92 \pm 0.01}$ & $\mathbf{0.92 \pm 0.00}$\\
MFI & $\mathbf{2.00 \pm 0.04}$ & 2.06 $\pm$ 0.04 & $\mathbf{8.76 \pm 0.31}$ & 9.52 $\pm$ 0.43 & $\mathbf{0.83 \pm 0.01}$ & 0.81 $\pm$ 0.01 \\
RHO & 1.44 $\pm$ 0.29 & $\mathbf{1.01 \pm 0.17}$  & 3.00 $\pm$ 0.99 & $\mathbf{1.62 \pm 0.43}$ & 0.66 $\pm$ 0.11 & $\mathbf{0.82 \pm 0.05}$ \\
ITW & $\mathbf{2.60 \pm 0.14}$ & 2.66 $\pm$ 0.11 & $\mathbf{16.69 \pm 1.36}$ & 17.42 $\pm$ 1.28 & $\mathbf{0.70 \pm 0.02}$ & 0.69 $\pm$ 0.02 \\
\bottomrule
\end{tabular}
    \label{tab:results_hoa}
\end{table}
\normalsize

\begin{table}[t!]
    \centering
    \caption{Performance of EPCN compared to our extension on different zeolites for Henry coefficient, Bold numbers indicate the best performance for a zeolite topology.}
\small
    \begin{tabular}{lcccccc}
\toprule
{} &  \multicolumn{2}{c}{MAE $\downarrow$} & \multicolumn{2}{c}{MSE $\downarrow$} & \multicolumn{2}{c}{R$^2$ $\uparrow$}\\
\cmidrule(lr){2-3}
\cmidrule(lr){4-5}
\cmidrule(lr){6-7}
{} &                  EPCN & this work & EPCN & this work & EPCN & this work\\
\midrule
MOR & $\mathbf{0.08 \pm 0.00}$ & $\mathbf{0.08 \pm 0.01}$ & $\mathbf{0.01 \pm 0.00}$ & $\mathbf{0.01 \pm 0.00}$  & $\mathbf{0.96 \pm 0.00}$ & 0.95 $\pm$ 0.01\\
MFI & $\mathbf{0.16 \pm 0.01}$ & $\mathbf{0.16 \pm 0.02}$ & $\mathbf{0.05 \pm 0.01}$ & $\mathbf{0.05 \pm 0.01}$ & $\mathbf{0.88 \pm 0.02}$ & 0.87 $\pm$ 0.02 \\
RHO & 0.22 $\pm$ 0.07 & $\mathbf{0.12 \pm 0.06}$  & 0.06 $\pm$ 0.03 & $\mathbf{0.03 \pm 0.02}$ & 0.48 $\pm$ 0.25 & $\mathbf{0.77 \pm 0.19}$ \\
ITW & $\mathbf{0.25 \pm 0.02}$ & 0.27 $\pm$ 0.06 & $\mathbf{0.10 \pm 0.01}$ & 0.13 $\pm$ 0.05 & $\mathbf{0.72 \pm 0.04}$ & 0.66 $\pm$ 0.14 \\
\bottomrule
\end{tabular}
    \label{tab:results_henry}
\end{table}
\normalsize

\begin{figure}[htb]
    \centering
    \includegraphics{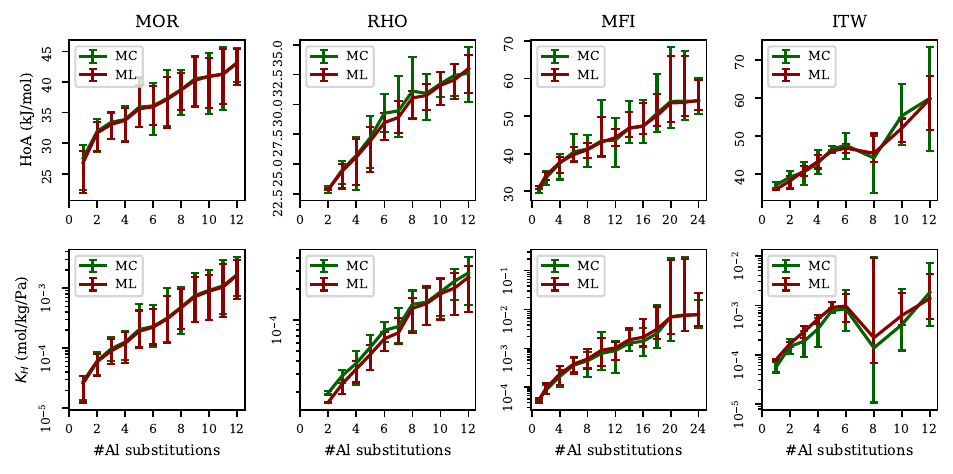}
    \caption{Heat of adsorption and Henry coefficient for different topologies predicted by MC and ML on the test set. Vertical bars indicate the 95\% confidence interval for each amount of substitutions.}
    \label{fig:MCvsml}
\end{figure}

\begin{figure}[htb]
    \centering
    \includegraphics{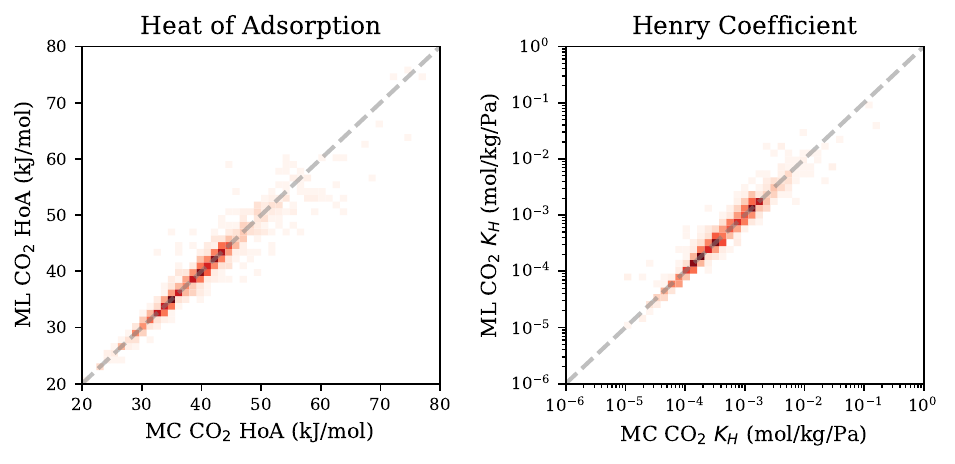}
    \caption{Heat of adsorption and Henry coefficient for all topologies predicted by MC and ML on the test set.}
    \label{fig:MCvsml_scatter}
\end{figure}

In Figure \ref{fig:MCvsml}, we compare the heat of adsorption distributions on the test set between our ML algorithm and MC simulations. For each topology, we took the best performing model. For all four topologies, the heat of adsorption obtained from the model seems to be in line with the heat of adsorption obtained from the simulations. It can be noticed that for all topologies, for some numbers of aluminium substitutions, the models underestimate the spread of the heat of adsorption and Henry coefficient distributions. This could be caused by a lack of similar examples in the training dataset, meaning that the model might not learn the effect of certain aluminum configurations. A similar trend can be seen in Figure \ref{fig:MCvsml_scatter}, where, especially for higher heat of adsorption values, the model seems to have a higher prediction error. Finally, there are significant differences in running time between the ML algorithm and MC simulations. Where MC simulations take hours to simulate the properties of a zeolite, the ML algorithm needs (milli)seconds to make a prediction. A detailed analysis of the running time of both algorithms can be found in the Supporting Information.  

\subsection{Interpretability}
\begin{figure}[hb!]
    \centering
    \includegraphics{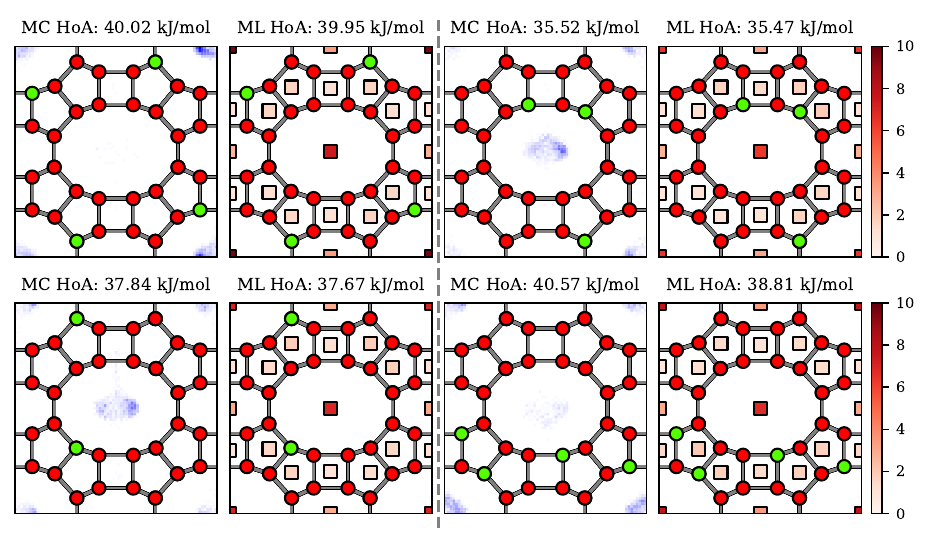}
    \caption{First and third column: CO$_2$ distribution in different Si/Al configurations of MOR. Second and fourth column: Final pore values used to predict the heat of adsorption. Green corresponds to Al atoms, while red corresponds to Si atoms. It can be seen that pores which do not adsorb any CO$_2$ have very low values, while pores which adsorb more CO$_2$ have (relatively) higher final embeddings.}
    \label{fig:co2emb}
\end{figure}

As a result of our proposed change in architecture, the model now predicts a scalar value per feature for each pore. In turn, the predictions can be interpreted as the contribution of each pore to the heat of adsorption/Henry coefficient. To qualitatively verify whether this is the case, we sampled 1500 CO$_2$ positions from the MC simulation for 4 structures. In Figure \ref{fig:co2emb}, we visualised the CO$_2$ distribution and the final pore values for these zeolites. It can be seen that the final values of the pore embeddings are proportional to the amount of times there was a CO$_2$ molecule present in the pore. For example, in the top left and bottom right cases, the CO$_2$ molecule is mainly present in the outer 12 ring. This is reflected in the final pore values, which in both cases is higher for the outer 12 ring. As such, the output of the model can be used to estimate which pores are responsible for the CO$_2$ adsorption.

\subsection{Inverse Design}
For each structure generated using our inverse design algorithm, we investigated whether it was part of the training set, and thus already seen by the model. Here, we also took into account all symmetry operations which leave the material invariant. Overall, we found that all structures generated for MFI were new, while 13.5\% of the structures generated for MOR were in the training set. Most of these structures where generated with a target heat of adsorption of up to 35 kJ/mol, and with 2 or 3 alumnium atoms. The possible number of unique MOR configurations with 2 to 3 aluminium atoms is relatively low, and is therefore mostly covered by the training set.

\begin{figure}[htb!]
     \centering
     \hspace*{\fill} 
     \begin{subfigure}[b]{0.485\textwidth}
         \centering
         \includegraphics{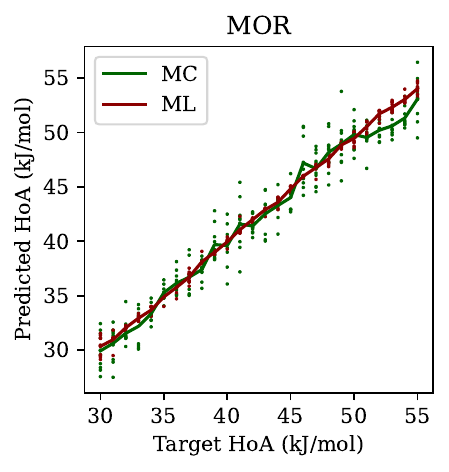}
         \label{fig:mor_gen}
     \end{subfigure}
     \hfill
     \begin{subfigure}[b]{0.485\textwidth}
         \centering
         \includegraphics{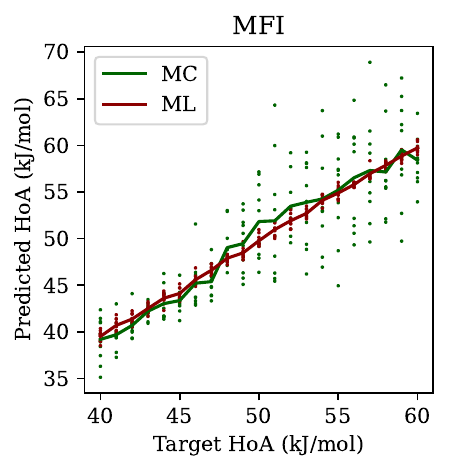}
         \label{fig:mfi_gen}
     \end{subfigure}
     \hspace*{\fill} 
     \caption{Predicted heat of adsorption against target heat of adsorption for structures generated using the genetic algorithm for MC and ML. The lines indicate the average predicted heat of adsorption for each target value, while individual dots indicate the heat of adsorption of the samples.}\label{fig:genstruct}
\end{figure}

Overall, the MAE between the ML and MC heat of adsorption values for MOR was 1.20, while the MSE was 2.42. For MFI, the MAE was 2.69, while the MSE was 13.24. For both materials, we see that there is a reasonable agreement between heat of adsorption predictions using ML and MC (Figure \ref{fig:genstruct}). For MOR, the ML model seems to slightly overestimate the heat of adsorption compared to MC, which is likely a result of the training set for MOR containing few examples with a heat of adsorption higher than 45 kJ/mol (Figure \ref{fig:dataset}). On the other hand, we see that for MFI the two methods are in agreement on average. However, especially for higher heat of adsorption values, we see that the ML model over- and underestimates these values compared to MC. A potential reason for this is the higher complexity of MFI compared to MOR, which in turn can pose a higher challenge for the model.

\section{Conclusions}
Overall, we have shown that ML can be a useful tool for modelling zeolites, with our proposed method achieving excellent performance on the heat of adsorption and Henry coefficients predictions, while also showing that this class of models could be interpretable. Our model can provide useful insights into the underlying mechanism that triggers CO$_2$ adsorption, such as the the contribution of each framework region into the heat of adsorption value. This strategy opens the possibility of identifying adsorbent features to maximize the performance of a target property. In addition, we have demonstrated that these models are a step towards inverse design of new materials, as it is possible to optimize structures for a target heat of adsorption using our model in combination with a genetic algorithm.

The main drawback of the model is that a new model needs to be trained for each zeolite topology. For zeolites, this is not a huge drawback, since the amount of Si/Al configurations is relatively large. To generalize this model, a potential solution could be to implement the parameter sharing idea based on composite building units (CBUs), since a single CBU can appear in more than one zeolite. As a result, the model architecture would retain additional expressiveness thanks to the idea of parameter sharing, while it can be used for multiple zeolite topologies at the same time.

The proposed method for the inverse design of zeolites introduced in this work is based on a discriminative ML model. Typically, discriminative models try to model the conditional probability distribution of a variable (heat of adsorption/Henry coefficient), given some input features (zeolite crystal structure). As such, they learn a direct mapping between the zeolite structure and the corresponding properties, which can result in the model not learning the joint distribution of the adsorption properties and zeolite configuration. On the other hand, generative models are able to model this joint probability. Therefore, this class of models could possibly provide a better framework for inverse design in the future, for example by extending methods such as Crystal Diffusion Variational Auto-Encoder \cite{xie2021crystal}.

In summary, the results from this work show the potential of our proposed method based on the EPCN architecture to predict and describe adsorption properties and materials features. We have shown that this is a step further in computational materials design, and this methodology can be extended to other properties and applications in materials science.

\section{Data and Software Availability}
The generated structures and their simulated properties are available on GitHub \cite{zpp2024}. The repository also contains the code for the Deep Learning models, experiments and results.

\section{Supporting Information}
\begin{itemize}
    \item Running time comparison between MC and ML (PDF)
\end{itemize}
\bibliography{mybib}

\end{document}